\RequirePackage[svgnames,dvipsnames,prologue]{xcolor}

\documentclass[authoryear,manuscript,acmlarge]{acmart}
\usepackage{qub}

\title{\qub{}: A Resource Aware Functional Programming Language}

\author{Apoorv Ingle}
\orcid{0000-0002-7399-9762}
\affiliation{
  \department{ACM ID: 7456710, Information and Telecommunication Technology Center}
  \institution{The University of Kansas}
  \country{USA}
}
\email{ani@ku.edu}

\renewcommand\footnotetextcopyrightpermission[1]{} 
\acmConference{}{}{}
\begin{document}

\maketitle
\thispagestyle{empty}

\section{Problem and Motivation}\label{sec:prob-and-motivation}\footnote{Extended abstract submitted to ICFP 2018 SRC. Research Advisor: J. Garrett Morris, garrett@ittc.ku.edu}
Managing resources---file handles, database connections, etc.---is a hard problem. Debugging resource leaks and runtime errors due to
resource mis-management are difficult in evolving production code. Programming languages with static type systems are great
tools to ensure erroneous code is detected at compile time. However, modern static type systems do little in the aspect of
resource management as resources are treated as normal values. We propose a type system, \qub{}, based on the logic
of bunched implications (\BI)\cite{ohearn_logic_1999} which models resources as first class citizens.
We distinguish two kinds of program objects---restricted and unrestricted---and two kinds of functions---sharing and separating.
Our approach guarantees resource correctness without compromising existing functional abstractions.

For a concrete example, we consider the case of file handling.
In Haskell, a file being closed twice or a file not being closed at all may cause run-time errors but it not flagged as a type error.
We represent separating functions, i.e. functions that do not share resources with their arguments using $\sepimp$, and sharing functions
i.e. functions that share resources with their arguments using $\shimp$. In \qub{}, the type signatures of the file handling API
explicitly states that they are separating in nature. This accounts for closing the file handle more than once.
Each program object needs to be explicitly dropped if it has to be treated as a resource,
as in linear type systems \citep{ahmed_l3_2007, mazurak_lightweight_2010, bernardy_linear_2017}. This accounts for failing to close the file handles.

Exception handling in Haskell can be done using \texttt{MonadError}\citep{liang_monad_1995}. However, it does not give a systematic
way of cleaning up resources in case of run-time exceptions. We consider the case where a critical section of the code
throws an exception as shown in \cref{fig:exception-handling-qub}. The \texttt{IOF} describes the fact that the computation
can throw exceptions, while \texttt{IO} does not. The \texttt{catch} function has a sharing
argument, hence it can access the file handle \texttt{fh} declared in the part of the code that can
throw exceptions and close it before exiting to prevent a memory leak.

\begin{figure}[h]
\begin{tabular}{c|c}
\begin{haskell}
openFile :: FilePath -* IO FileHandle
closeFile :: FileHandle -* IO ()
readFile :: FileHandle
         -* IOF (String, FileHandle)
writeFile :: String
          -* FileHandle
          -* IOF ((), FileHandle)

throw :: Exception -* IO a
catch :: IOF a -* (Exception -* IO a)
      ->> IO a
\end{haskell}
&
\begin{haskell}
readFromFile :: FilePath
             -* IO (Either String String)
readFromFile fpath =
do fh <- openFile fpath
   ((s, fh) <- readLine fh
   let l = caps s
   closeFile fh
   return \dollar Right l)
       `catch` (\e -> do closeFile fh
                         return \dollar
                            Left "read file error")
\end{haskell}
\end{tabular}
\caption{File and Exception Handling in \qub{}}
\label{fig:exception-handling-qub}
\end{figure}

\section{Background and Related Work}\label{sec:backgrond}
Type systems based on linear logic\citep{girard_linear_1987, wadler_taste_1993, ahmed_l3_2007, mazurak_lightweight_2010, bernardy_linear_2017} provide one technique
to solve the resource control problem. They restrict the structural rules of weakening and contraction
to view all values as resources. This changes the meaning of the connectives as well.
Linear implication $A \rightspoon B$ means ``A is consumed to obtain B''.
We also get additive and multiplicative fragments of conjunction ($A \otimes B$ means ``both A and B'' and $A \with B$ means ``choose between A and B'').
There is, however, an awkward asymmetry in this system---while $\rightspoon$ is the right adjoint of $\otimes$, $\with$ has no such counterpart.
Logic of \BI\citep{pym_semantics_2002} repairs this asymmetry between implication and conjunction.
It uses trees as contexts, where the internal nodes are either comma ($,$) or semicolon ($;$) and leaf nodes are the propositions.
The structural rules---weakening and contraction---are prohibited for propositions connected using ($,$).
$\Gamma;\Delta \vdash \Gamma$ but $\Gamma,\Delta \nvdash \Gamma$.
The multiplicative conjunction $\otimes$ gets a multiplicative implication $\sepimp$
and the additive conjunction $\with$ gets the additive implication $\shimp$ as its right adjoint. The Curry-Howard interpretation of \BI is
in terms of sharing in rather than linear logic's consumption. If the function does not share resources with its argument $\sepimp$
is used, while if the function shares resources with its arguments, $\shimp$ is used instead.

Jones\citep{jones_theory_1994, jones_qualified_2003} introduces qualified types, a general framework to incorporate predicates for polymorphism.
The Hindley-Milner type system\citep{milner_theory_1978} extended with qualified types\citep{jones_simplifying_1995}
can express type classes with functional dependencies\citep{mark_type_2000}, and first class polymorphism\citep{jones_first-class_1997}.
Morris\citep{morris_best_2016} uses qualified types to design Quill, a functional language with linear calculus. In Quill, the predicate $\Un{\tau}$ specifies
the type $\tau$ is unrestricted i.e. it can be duplicated or dropped at will, or it does not contain any resources. Proof theoretically, the type is tagged
unrestricted whenever weakening and contraction is admissible. A binary predicate $\geq$ helps generalize function definition in presence of
restricted types. $\tau \geq \tau'$ specifies that type $\tau$ admits more structural rules than type $\tau'$.

\section{Approach and Uniqueness}\label{sec:approach}
\qub{} is an extension of standard call-by-name lambda calculus based on logic of \BI.
We introduce two kinds of lambdas associated with the two implications.
$\lambda^{\sepimp}x.M$ introduces a separating function $\sepimp$, while $\lambda^{\shimp} x. M$
introduces a sharing arrow $\shimp$. We generalize the use of trees as contexts in \BI to graphs of sharing information.
We represent sharing graphs as adjacency lists in the environment context. A triple $(x^{\vec{y}}:\tau) \in \Gamma$
would mean $x$ of type $\tau$ is in sharing with $\vec{y}$. The sharing relation is a symmetric, reflexive and non-transitive.
We say that the contexts are in complete sharing---$\Gamma \oplus \Delta$---if all the variables are shared and they
are disjoint---$\Gamma \circledast \Delta$---if they are not shared. We formally define them in \cref{fig:aux-functions}, where $\mathbin{\#}$ means disjoint.
The predicates $\ShFun{\phi}$ and $\SeFun{\phi}$ range over sharing and separating functions respectively.
We include predicates $\Un{\tau}$ and $\tau \geq \tau'$ as is from Quill. The complete type system is shown in \cref{fig:type-system}.
{\small
\begin{figure}[h]\centering
    \begin{minipage}[h]{0.45\linewidth}
    \begin{flalign*}
      \texttt{Vars}(\Gamma, x^{\vec{y}}) &= \texttt{Vars}(\Gamma) \cup \{ x \}\\[-5pt]
      \texttt{Shared}(\Gamma, x^{\vec{y}}) &= \texttt{Shared}(\Gamma) \cup \{ \vec{y} \}\\[-5pt]
      \texttt{Used}(\Gamma) &= \texttt{Vars}(\Gamma) \cup \texttt{Shared}(\Gamma)
      \end{flalign*}
    \end{minipage}%
    \begin{minipage}[ht]{0.45\linewidth}
      \begin{flalign*}
        (\Gamma, x^{\vec{y}})^{[a \mapsto \vec{b}]} &=
        \begin{cases}
          a \notin \vec{y}\ \ \ \ (\Gamma^{[a \mapsto \vec{b}]}, x^{\vec{y}}:\tau)\\
          a \in \vec{y}\ \ \ \  (\Gamma^{[a \mapsto \vec{b}]}, x^{(\vec{y}\backslash a)\cup\vec{b}}:\tau)
        \end{cases}\\[-5pt]
      \Gamma^{[\vec{a} \mapsto \vec{b}]} &= (\dots((\Gamma^{[a_1 \mapsto \vec{b}]})^{[a_2 \mapsto \vec{b}]})^{\dots})^{[a_n \mapsto \vec{b}]}
      \end{flalign*}
    \end{minipage}\\[-5pt]
    \begin{minipage}[ht]{0.45\linewidth}
      \begin{flalign*}
      \Gamma \circledast \Gamma' &= \Gamma \sqcup \Gamma'\ \ \
           \textit{if}\ \texttt{Vars}(\Gamma) \mathbin{\#} \texttt{Used}(\Gamma') \wedge \texttt{Vars}(\Gamma') \mathbin{\#} \texttt{Used}(\Gamma) \\[-5pt]
      \Gamma \oplus \Gamma'   &= \Gamma \sqcup \Gamma'\ \ \ \textit{if}\ \texttt{Used}(\Gamma) = \texttt{Used}(\Gamma')
    \end{flalign*}
  \end{minipage}
  \caption{Auxiliary Functions}
  \label{fig:aux-functions}
\end{figure}
\begin{figure}[h]\centering
    \begin{minipage}{.35\textwidth}
      \begin{prooftree}
        \AxiomC{{\color{white}$\Gamma \circledast \Delta \circledast$}} \RightLabel{[ID]}
        \UnaryInfC{$P \mid x^{\vec{y}} : \sigma \vdash x : \sigma $}
      \end{prooftree}
    \end{minipage}%
    \begin{minipage}{.50\textwidth}
      \begin{prooftree}
        \AxiomC{$P \mid \Gamma \circledast \Delta \circledast \Delta \vdash M : \sigma$}
        \AxiomC{$P \vdash \Delta\ \texttt{un}$} \RightLabel{[CTR-UN]}
        \BinaryInfC{$P \mid \Gamma \circledast \Delta \vdash M : \sigma$}
      \end{prooftree}
    \end{minipage}\\[3pt]

    \begin{minipage}{.35\textwidth}
      \begin{prooftree}
        \AxiomC{$P \mid \Gamma \oplus \Delta \oplus \Delta\vdash M : \sigma$}\RightLabel{[CTR-SH]}
        \UnaryInfC{$P \mid \Gamma \oplus \Delta \vdash M : \sigma$}
      \end{prooftree}
    \end{minipage}%
    \begin{minipage}{.50\textwidth}
      \begin{prooftree}
        \AxiomC{$P \mid \Gamma \vdash M : \sigma$}
        \AxiomC{$P \vdash \Delta\ \texttt{un}$} \RightLabel{[WKN-UN]}
        \BinaryInfC{$P \mid \Gamma \circledast \Delta \vdash M : \sigma$}
      \end{prooftree}
    \end{minipage}\\[3pt]

    \begin{minipage}{.35\textwidth}
      \begin{prooftree}
        \AxiomC{$P  \mid \Gamma \vdash M : \sigma$} \RightLabel{[WKN-SH]}
        \UnaryInfC{$P \mid \Gamma \oplus \Delta \vdash M : \sigma$}
      \end{prooftree}
    \end{minipage}%
    \begin{minipage}{0.60\textwidth}
      \begin{prooftree}
        \AxiomC{$P \mid \Gamma \vdash M : \sigma$}
        \AxiomC{$P' \mid \Gamma'_{x} \sqcup x: \sigma \vdash N: \tau$} \RightLabel{[LET]}
        \BinaryInfC{$P \cup P' \mid \Gamma \sqcup \Gamma' \vdash (\Let{x}{M}{N}): \tau$}
      \end{prooftree}
    \end{minipage}\\[3pt]

    \begin{minipage}{0.45\textwidth}
      \begin{prooftree}
        \AxiomC{$P \mid \Gamma \vdash M: \sigma$}
        \AxiomC{$t \notin \texttt{fvs}(\Gamma) \cup \texttt{fvs}(P)$}\RightLabel{[$\forall$ I]}
        \BinaryInfC{$P \mid \Gamma \vdash M: \forall t. \sigma$}
      \end{prooftree}
    \end{minipage}%
    \begin{minipage}{0.45\textwidth}
      \begin{prooftree}
        \AxiomC{$P \mid \Gamma \vdash M: \forall t.\sigma$}\RightLabel{[$\forall$ E]}
        \UnaryInfC{$P \mid \Gamma \vdash M: [\tau \backslash t] \sigma $}
      \end{prooftree}
    \end{minipage}\\[3pt]

    \begin{minipage}{0.45\textwidth}
      \begin{prooftree}
        \AxiomC{$P, \pi \mid \Gamma \vdash M : \rho$} \RightLabel{[$\Rightarrow$ I]}
        \UnaryInfC{$P \mid \Gamma \vdash M : \pi \Rightarrow \rho$}
      \end{prooftree}
    \end{minipage}%
    \begin{minipage}{0.45\textwidth}
      \begin{prooftree}
        \AxiomC{$P \mid \Gamma \vdash M : \pi \Rightarrow \rho$}
        \AxiomC{$P \vdash \pi$} \RightLabel{[$\Rightarrow$ E]}
        \BinaryInfC{$P \mid \Gamma \vdash M: \rho$}
      \end{prooftree}
    \end{minipage}\\[3pt]

    \begin{minipage}{0.35\textwidth}
      \begin{prooftree}
        \AxiomC{$P \Rightarrow \texttt{ShFun}\ \phi\ \ \ \ \
          P \vdash \Gamma \geq \phi$}\noLine\def\extraVskip{-0.2pt}
        \UnaryInfC{$P \mid \Gamma^{[\emptyset\mapsto \{x\}]},x^{\text{Vars}(\Gamma)}: \tau \vdash M : \tau'$}\RightLabel{[$\shimp$ I]}\def\extraVskip{2pt}
        \UnaryInfC{$P \mid \Gamma \vdash \lambda^{\shimp}x. M : \phi \tau \tau'$}
      \end{prooftree}
    \end{minipage}%
    \begin{minipage}{0.55\textwidth}
      \begin{prooftree}
        \AxiomC{$P \Rightarrow \texttt{ShFun}\ \phi$}\noLine\def\extraVskip{0pt}
        \UnaryInfC{$P \mid \Gamma \vdash M : \phi \tau \tau'\ \ \ \ \
          P \mid \Delta \vdash N : \tau'$} \RightLabel{[$\shimp$ E]}\def\extraVskip{2pt}
        \UnaryInfC{$P \mid \Gamma \oplus \Delta \vdash M N : \tau'$}
      \end{prooftree}
    \end{minipage}\\[3pt]

    \begin{minipage}{0.35\textwidth}
      \begin{prooftree}
        \AxiomC{$P \Rightarrow \texttt{SeFun}\ \phi\ \ \ \ \ P \vdash \Gamma \geq \phi$}\noLine\def\extraVskip{0pt}
        \UnaryInfC{${\color{white}\ \ \ \ }P \mid \Gamma,x^{\emptyset}: \tau \vdash M : \tau'{\color{white}\ \ \ \ }$} \RightLabel{[$\sepimp$ I]}\def\extraVskip{2pt}
        \UnaryInfC{$P \mid \Gamma \vdash \lambda^{\sepimp}x. M : \phi \tau \tau'$}
      \end{prooftree}
    \end{minipage}%
    \begin{minipage}{0.55\textwidth}
      \begin{prooftree}
        \AxiomC{$P \Rightarrow \texttt{SeFun}\ \phi$}\noLine\def\extraVskip{0pt}
        \UnaryInfC{$P \mid \Gamma \vdash M : \phi \tau \tau'\ \ \ \ P \mid \Delta \vdash N : \tau$}\RightLabel{[$\sepimp$ E]}\def\extraVskip{2pt}
        \UnaryInfC{$P \mid \Gamma \circledast \Delta \vdash M N : \tau'$}
      \end{prooftree}
    \end{minipage}

  \caption{\qub{} Type System}
  \label{fig:type-system}
\end{figure}
}
\section{Results and Contributions}\label{result}
\qub{} is a novel sub-structural $\lambda$-calculus that generalizes Curry-Howard Interpretation of \BI. We have developed a sound and complete
syntax directed \qub{} type system and designed a type inference algorithm based on Algorithm $\M$\citep{lee_proofs_1998}.
We have extended our system to support kinds with user defined type constructors allowing programmers to define data types with sharing and separating fields.
The use of monads with sharing and separating functions can statically detect resource errors, while expressing patterns like exceptions and non-determinism
that are difficult to capture in linear languages as described in previous section.

\newpage

\bibliographystyle{ACM-Reference-Format}
\bibliography{qub-icfp-2018}

\end{document}